\documentclass[11pt,a4paper]{article}

\title{Efficient computational homogenisation of 2D beams of heterogeneous elasticity using the patch scheme}

\author{%
Thien Tran-Duc\thanks{%
Mathematical Sciences, University of Adelaide,  South Australia.
\protect\url{mailto:thien.tran-duc@adelaide.edu.au},
\protect\url{https://orcid.org/0000-0002-2004-5156}}
\and 
J.E. Bunder\thanks{%
UniSA STEM, University of South Australia, Australia.
\protect\url{mailto:Judy.Bunder@unisa.edu.au},
\protect\url{https://orcid.org/0000-0001-5355-2288}}
\and 
A.J. Roberts\thanks{%
University of Adelaide,  South Australia.
\protect\url{mailto:ProfAJRoberts@protonmail.com},
\protect\url{https://orcid.org/0000-0001-8930-1552}}
}

\IfFileExists{ajr.sty}{%
\usepackage[a5paper,margin=10mm,tmargin=15mm]{geometry} 
}{%
}

\let\MakeUppercase\sf
\usepackage{amsmath,amssymb,microtype}
\usepackage[defaultlines=3,all]{nowidow}

\usepackage[leftcaption]{sidecap}

\usepackage[backend=bibtex8 %
    ,style=authoryear %
    ,backref=true %
    ]{biblatex}
\IfFileExists{ajr.sty}
  {\bibliography{ajr,bib,bibexport}}
  {\bibliography{bibexport}}
\AtEndDocument{{\raggedright\printbibliography}}
\let\citet\textcite
\let\citep\parencite
\makeatletter
\def\cite{\@ifnextchar[{\parencite}{\textcite}}%
\makeatother
\DeclareFieldFormat{url}{%
  \iffieldundef{doi}{%
  \url{#1}}{}}
\DeclareFieldFormat{urldate}{%
  \iffieldundef{doi}{%
  \mkbibparens{\bibstring{urlseen}\space#1}}{}}

\usepackage{tikz,pgfplots}
\pgfplotsset{compat=newest}
\def\tikzsetnextfilename#1{} %
\usepgfplotslibrary{external}
\usepackage{doi}
\hypersetup{
    colorlinks   = true, %
    urlcolor     = blue, %
    linkcolor    = blue, %
    citecolor   = magenta,
    } 
\usepackage[nameinlink,capitalise,noabbrev]{cleveref}
\crefname{equation}{}{}
\let\ref\cref
\let\eqref\cref
\let\autoref\cref

\usepackage[colorinlistoftodos,textsize=footnotesize]{todonotes}
  \setuptodonotes{inline,color=orange!20}
  \ifx\undefined
     
  \else
     
  \fi

\def\RaisedName#1{}%
\def\ajrSplit#1#2\ajrEndSplit{\def\ajrcar{#1}\def\ajrcdr{#2}}
\def\Vec{}
\renewcommand{\Vec}[1]{%
  \typeout{**** Command #1v denotes vector #1}%
  \expandafter\ajrSplit#1\ajrEndSplit%
  \ifx\ajrcdr\empty %
    \expandafter\DeclareRobustCommand\csname#1v\endcsname%
    {{\RaisedName{\ensuremath{\backslash}#1v}%
      \ensuremath{\ifx#1i \vec\imath
        \else\ifx#1j \vec\jmath
        \else\vec #1\fi\fi}%
      \global\csname UsedVec#1true\endcsname%
    }}%
  \else %
    \expandafter\DeclareRobustCommand\csname#1v\endcsname%
    {{\ensuremath{\vec{\csname#1\endcsname}}}%
      \global\csname UsedVec#1true\endcsname}%
  \fi%
  \expandafter\newif\csname ifUsedVec#1\endcsname
  \csname UsedVec#1false\endcsname
  \AtEndDocument{\csname ifUsedVec#1\endcsname\else
  \typeout{**** Vec symbol #1v not used.}
  \fi}
  }
\newcommand{\Cal}[1]{%
  \typeout{**** Command c#1 denotes calligraphic #1}%
  \expandafter\DeclareRobustCommand\csname c#1\endcsname%
  {{\RaisedName{\detokenize{\c}\hspace{-0.5em}#1}%
  \ensuremath{\mathcal #1}%
  \global\csname UsedCal#1true\endcsname  %
  }}%
  \expandafter\newif\csname ifUsedCal#1\endcsname
  \csname UsedCal#1false\endcsname
  \AtEndDocument{\csname ifUsedCal#1\endcsname\else
  \typeout{**** Cal symbol c#1 not used.}
  \fi}
  }

\renewcommand{\vec}[1]{\text{\boldmath$#1$}}
\newcommand{\dx}{\delta_x}
\newcommand{\dy}{\delta_y}
\newcommand{\E}{\cdot 10^}
\def\half{\frac12}

\def\matlab{\textsc{Matlab}}
\def\lhk{\textsc{lhk}}
\def\pde{\textsc{pde}}
\def\ode{\textsc{ode}}

\def\rve{\textsc{rve}}
\Vec e\Vec f\Vec k\Vec n\Vec q\Vec x\Vec u
\Vec{varepsilon}\Vec{sigma}\Vec I\Vec J\Vec F
\Cal U

\newcommand{\D}[2]{\mathchoice
  {\frac{\partial #2}{\partial #1}}%
  {{\partial #2}/{\partial #1}}%
  {{\partial #2}/{\partial #1}}%
  {{\partial #2}/{\partial #1}}%
  }
\newcommand{\Dn}[3]{\mathchoice
  {\frac{\partial^{#2} #3}{\partial #1^{#2}}}%
  {{\partial^{#2} #3}/{\partial #1^{#2}}}%
  {{\partial^{#2} #3}/{\partial #1^{#2}}}%
  {{\partial^{#2} #3}/{\partial #1^{#2}}}%
  }
\newcommand{\DD}[2]{\Dn{#1}2{#2}}

\newcommand{\Ord}[1]{\ensuremath{\mathcal O%
  \mathchoice{\big(#1\big)}{\big(#1\big)}{(#1)}{(#1)}%
  }}

\newcommand{\I}{\operatorname{i}}
\def\E#1{\textsc{e}\ifx#1-{-}\else\ifx#1+{+}\else#1\fi\fi}

\def\temp#1{#1}%
\def\oSym{\temp{$\color{green!70!black}\circledcirc$}}
\def\xSym{\temp{$\color{green!70!black}\otimes$}}
\def\uSym{\temp{$\color{blue}\blacktriangleright$}}
\def\vSym{\temp{$\color{red}\blacktriangle$}}

\IfFileExists{ajr.sty}%
   {\let\inPlot\input}%
   {\let\inPlot\includegraphics}

\usepackage{versions}
\renewenvironment{comment}{\color{red!50!black}\small\sf}{}
\excludeversion{comment}%

\begin{document}

\maketitle

\begin{abstract}
Modern `smart' materials have complex heterogeneous microscale structure, often with unknown macroscale closure but one we need to realise for large scale engineering and science.  
The multiscale Equation-Free Patch Scheme empowers us to non-intrusively, efficiently, and accurately predict the large scale, system level, solutions through computations on only small sparse patches of the given detailed microscale system.  
Here the microscale system is that of a 2D beam of heterogeneous elasticity, with either fixed fixed, fixed-free, or periodic boundary conditions.  
We demonstrate that the described multiscale Patch Scheme simply, efficiently, and stably predicts the beam's macroscale, with a controllable accuracy, at finite scale separation.
Dynamical systems theory supports the scheme.  
This article points the way for others to use this systematic non-intrusive approach, via a developing toolbox of functions, to model and compute accurately macroscale system-levels of general complex physical and engineering systems.  

\end{abstract}

\tableofcontents

\section{Introduction}

With advances in materials science and engineering, it is now possible to engineer materials with desirable mechanical strengths and physical and chemical properties for specific applications \citep[e.g.,][]{Torquato2010, Somnic2022}. 
Metal alloy, polymer-based materials, lattice materials, and composite materials are common examples of such materials. 
Fabricated at small length scales from multiple constituents, these materials are highly heterogeneous and hence it is non-trivial to characterise their macroscale mechanical properties. 
A significant impediment is that extant accurate mathematical models of heterogeneous materials are typically of microscale nature, and full computational solutions are thus prohibitively expensive. 
Here we circumvent this obstacle by developing an efficient, accurate and stable multiscale computational homogenisation of heterogeneous beams.

Beams made from innovative heterogeneous or composite elastic materials need to be understood and characterised, and hence continue to be extensively studied \citep[e.g.,][]{Pau2022, Klarmann2019, Rupprecht2016}.  
This article develops applying the the Equation-Free Patch Scheme \citep[e.g.,][]{Kevrekidis2003, Kevrekidis2004, Samaey2005, Samaey2006}, sometimes called the gap-tooth scheme, to predict the dynamics of 2D heterogeneous visco-elastic beams.
One important characteristic of the patch scheme is that it is \emph{non-intrusive}---it just `wraps around' whatever microscale code is trusted by a user (\cref{secSnifw}).
Here, \cref{sec:microscalemodel,sec:discretisation} specify the microscale lattice model of heterogeneous 2D visco-elasticity trusted to accurately model interesting beam dynamics \citep[akin to][]{Virieux84, Virieux86}. 
Importantly, the patch scheme provides a macroscale computational homogenisation \emph{without} requiring any variational principle, and so the scheme straightforwardly applies to visco-elasticity (as herein) and \text{to nonlinearity (future work).}

Broadly, homogenisation of heterogeneous elasticity is performed with three different classes of methods \cite[e.g.,][]{Fish2021}.
Firstly, analytic or algebraic homogenisation analyses Representative Volume Elements~(\rve{}s) to derive algebraic differential equations for the macroscale averaged deformations \cite[e.g.,][]{RamirezTorres2018, Pau2022}.
For a beam, let \(U(x,t)\)~denote either the macroscale `averaged' longitudinal (compression) displacement or the lateral (bending) displacement, then the derived macroscale model is a \pde\ such as \(\DD tU=A_2\DD xU\), or higher-order analysis gives a \pde\ like \begin{equation}
\DD tU=A_2\DD xU+A_4\Dn x4U+\cdots\,,
\label{EqHomoPDE}
\end{equation}
for some homogenised coefficients~\(A_k\) \cite[e.g.,][]{Roberts2013a}.
This methodology typically introduces an~\(\epsilon\) that measure the spatial scale separation between macro- and micro-scales, and are typically derived via multiple scales theory that invokes the limit \(\epsilon\to0\).
Secondly, upscaling methods like numerical homogenisation\slash reduced-order-modelling similarly derive macroscale generalised `stress-strain' closure relationships implicit in \pde{}s like~\cref{EqHomoPDE} but based upon the precomputed numerical compilation of a `library' of canonical deformations \cite[e.g.,][]{Kouznetsova2001, Geers2010, Geers2017, Raju2021, Schneider2021}.
The theory underlying such methodology also assumes a scale separation parameter~\(\epsilon\to0\) (albeit often implicitly).
Thirdly, resolved scale, computational homogenisation methods like the Heterogeneous Multiscale Method \cite[e.g.,][]{Abdulle2012} and the Patch Scheme compute on-the-fly necessary macroscale information from local detailed microscale simulations---there is no preprocessing.
Such methods do not assume an algebraic closure; instead an effective closure is provided by the local microscale simulations.
The distinguishing feature of the patch scheme developed herein is that the underlying theory \cite[e.g.,][]{Roberts06d, Roberts2011a, Bunder2020a} presumes no~\(\epsilon\), no~limit as \(\epsilon\to0\), and instead applies rigorously at the \emph{finite} scale separation of real physical scenarios.
This feature is important, as in many practical beam modelling scenarios the scale separation parameter may be as large as \(0.1\)--\(0.3\) (e.g., see the gaps in the spectra of \cref{fig:Inhomobeam_EigVal,fig:Homobeam_EigVal,fig:Inclusionbeam_EigVal}).
Further, theory \cite[e.g.,][]{Bunder2020a} and the examples of \cref{SSfflb,SSpcpn} show that the patch scheme has controllable accuracy at finite scale separation, even up to eight significant digits in the macroscale predictions.

\begin{comment}
In materials with highly random heterogeneity and non-linear microscale dynamics, it is often difficult to construct constitutive models using conventional homogenisation techniques \citep{Geers2017}.  
Other common issues related to homogenisation are questions concerning the convergence of the truncated power series for finite\slash large deformations and expensive computational costs  \citep{Somnic2022}.
If constitutive models cannot be obtained, direct simulations for microscale dynamics are preferred. \cite{Kouznetsova2001} adopted a micro-macro modelling technique in which the microscale dynamics within \rve{}s are explicitly modelled and simulated at macroscale grid nodes.  
The resulting averaged stress tensor in the \rve{}s is then adopted for macroscale simulations on the macroscale grid. 
This explicitly microscale-simulated technique has been used to successfully solve many problems \citep{Fleischhauer2016, Raju2021}, although it is quite expensive due to the concurrent micro-macro computations.

Equation-free modelling is a computational technique for simulating multiscale systems \citep{Kevrekidis2003, Kevrekidis2004, Samaey2005, Samaey2006}.  
It has similarities to the micro-macro homogenization technique in that it solves explicit microscale simulations, however equation-free modelling does not derive or solve any macroscale governing equations.  
Instead, macroscale quantities are calculated directly from the microscale simulations. 
\end{comment}

The multiscale Patch Scheme achieves high computational efficiency by only computing the microscale complex physical phenomena in small, sparse, patches of space (more generally in small, sparse, patches of \emph{space-time}, e.g., \cite{Kevrekidis2004}).
Patches are analogous to the \rve{}s of other methodologies.
The patches are positioned in space to resolve the desired macroscale phenomena. 
\cref{sec:coupling} discusses how patches are coupled via inter-patch interpolation which determines field values on the patch edges.
In essence, the microscale simulations within a patch automatically provide a `high-order' accurate homogenised closure on the scale of the macroscale grid of patches.
Thus the patch scheme provides a computationally efficient homogenisation (\cref{sec:coupling,SScwsom}).
Importantly, theory \citep{Bunder2020a} proves the following: 
there is no need for debatable assumed boundary conditions on \rve{}s;  
as discussed by \cref{SSpsemd}, there is no need for presumed macroscale averages;
there is no need for arbitrary averaging in that there is no oversampling regions, no buffer regions, no action regions; and lastly, 
there is no need to guess specific fast/slow variables (\cref{SSpsemd}).
The patch scheme is much \text{more physically interpretable.}

\begin{comment}In a recent review for \textsc{nasa}, \cite{NASA2018} discussed the accuracy and adaptability of available multiscale methods and concluded that there is a ?Lack of useful automatic methods for linking models and passing information between scales?. Our equation-free computational schemes fill this lack by using a given microscale model directly, with no simplification or trans- formation (a major difference with many other schemes, which may, for example, require such modifications to explicitly separate ?slow? and ?fast? components [14]), and invoking generically crafted coupling conditions to ensure macroscale accuracy. The methodology bypasses the derivation of macroscopic evolution equations through computing the microscale only on small, well-separated, patches in space.\end{comment}

In their recent review of multiscale modelling in materials, \cite{Fish2021} [p.774] comment that pursuing ``a multiscale approach involves a trade-off between increased model fidelity with the added complexity, and corresponding reduction in precision and increase in uncertainty''.
With the multiscale patch scheme there is no need for such a trade-off.
Further, \cite{Fish2021} [p.781] assert that ``Progress in this and similar methods requires research on accurate and efficient lifting operators for specific applications.''
Theory underlying the patch scheme proves \citep{Bunder2020a}, and the elastic beam application developed herein confirms, that the patch coupling of \cref{sec:coupling} in essence provides such ``accurate and efficient lifting'' \text{in general circumstances.}

\begin{comment}
There are several major advantages of equation-free modelling compared to  homogenisation and its many variants. 
Firstly, there is no need for constitutive models of the microscale dynamics. 
Secondly, simulations are carried out only within very small patches, not across the entire macroscale domain, and hence computation costs are relatively inexpensive. 
Thirdly, the macroscale dynamics of the heterogeneous system are directly calculated from the simulated patches without needing to solve any additional macroscale equations. 

The two main components of the equation-free multiscale patch simulation are the microscale simulations in small patches and  the patch coupling.  
In the patch simulations, any suitable microscale simulator can be used, for example, molecular dynamics, finite difference or finite element methods, or Lagrangian particle-based methods \citep{Alotaibi2018}. 
\cref{sec:coupling} describes patch coupling, which is performed using either spectral or polynomial  interpolation  \cite[]{Bunder2020a} and constrains the fields values along the edges of each patch. 
\end{comment}

The patch coupling conditions are distinct and unrelated to the physical boundary conditions on the macroscale domain of the system. 
Here the patches extend across the beam, so physical stress-free conditions are applied at the top and bottom of the beam.
\cref{sec:load} explores examples of loaded heterogeneous beams with fixed-fixed end boundary conditions, and beams with fixed-free ends.
\cref{Sec:Heterobeam} explores the accuracy and stability of heterogeneous beams that are macroscale periodic.
In all cases, the patch edges \emph{not} on a physical boundary have values set by the inter-patch interpolation of \cref{sec:coupling}.

Most research and development has been done on homogenisation for problems seeking either equilibria, quasi-static laws, or a Helmholtz equation for strictly periodic linear oscillations \cite[e.g.,][]{Fish2021, Schneider2021, Raju2021}.
As does \cref{sec:load}.
However, addressing dynamics is a much more stringent test of a multiscale scheme.
\cref{Sec:Heterobeam} discusses how a multiscale scheme may be perfectly adequate for such `equilibria' problems, yet be ruinous for simulating and predicting dynamics in time.
By design \citep{Bunder2020a}, the patch scheme we implement here automatically preserves important symmetries present in the underlying microscale dynamics, and so provides stable predictive dynamical simulations (\cref{sec:eg,Sec:SoftInclusion}), even \text{without damping (\cref{SSeswd}).}

Recall that the equation-free patch scheme is designed to wrap around any given microscale system to then non-intrusively empower efficient and accurate macroscale prediction.
A freely available Equation-Free Toolbox for \matlab\ or Octave \citep{Maclean2020a, Roberts2019b} invokes a user-defined function of the microscale physics, and applies the patch scheme as a `black-box'.
A user also specifies the macroscale geometry and boundary conditions of interest in space-time.
The macroscale results of \cref{sec:load,Sec:Heterobeam} are all obtained by  applying functions of the Equation-Free Toolbox on the given microscale visco-elasticity \text{of \cref{sec:discretisation}.}

\section{Patch scheme for 2D elastic beam}
\label{Sec:Patchscheme}

Building on an earlier pilot study \citep{Roberts2023a}, this section introduces one way to design a multiscale patch scheme to empower efficient and accurate macroscale predictions of visco-elastic beams.

\subsection{Elastic stresses and equations of motion} 
\label{sec:microscalemodel}
In an elastic material the strain tensor
     \(\varepsilon(\xv,t)   := \frac{1}{2}\left[\nabla \uv +(\nabla \uv )^T\right]\), 
in which $\uv(\xv,t) :=(u,v)$ is the displacement vector in directions~\(x,y\) respectively.
By this definition, the strain tensor~$\varepsilon$ is symmetric.
Elastic deformations result in internal elastic stresses 
    \(\sigma := 2\mu\varepsilon +\lambda \operatorname{Tr}(\varepsilon )I\) ,
in which~$\lambda$ and~$\mu$ are Lam\'e constants that characterise the elasticity of the material.
Here we focus on an elastic beam in a 2D \(xy\)-space.
The stress tensor is thus explicitly 
\begin{subequations}\label{eqs:stressDef}%
\begin{align}
\sigma_{xx} &= 2\mu \varepsilon_{xx} +\lambda
(\varepsilon_{xx} +\varepsilon_{yy}) 
= (2\mu+\lambda)\D xu + \lambda \D yv,
\\
\sigma_{yy} &= 2\mu \varepsilon_{yy} +\lambda
(\varepsilon_{xx} +\varepsilon_{yy})
=\lambda \D xu + (2\mu+\lambda)\D yv,
\\
\sigma_{xy} &=\sigma_{yx} = 2\mu \varepsilon_{xy} 
            =\mu \left(\D yu + \D xv\right).    
\end{align}
\end{subequations}
Stress variations inside the material then drive motion of the beam.
Via continuum modelling of Newton's second law, the displacement vector evolves according to
\begin{equation}
    \rho\DD t\uv=\boldsymbol{\nabla} \cdot\sigma +\fv  \,,\label{E_Disp}
\end{equation} 
in which $\boldsymbol{\nabla}\cdot\sigma$~is the elastic force density arising from beam deformations, and $\fv $~is the external force density acting on the beam. 
By introducing non-dimensional quantities, $\uv^*=\uv/L_0$, $t^*=t/t_0$, $\boldsymbol{\nabla}^*=L_0\boldsymbol{\nabla}$, $\sigma^*=\sigma/\sigma_0$ and $\fv ^*=\fv /f_0$, and then dividing both sides by~$\rho L_0/t_0^2$, the non-dimensionlized equation of beam motion is 
\begin{equation*}
    \DD {{t^*}}{\uv^*}=\frac{\sigma_0t_0^2}{L_0^2\rho}\boldsymbol{\nabla}^* \cdot\sigma^* +\frac{f_0t_0^2}{L_0\rho}\fv ^* .%
\end{equation*}
Then, on choosing $L_0=L/2\pi$, $t_0=L_0\sqrt{E/\rho}$, $\sigma_0=L_0^2\rho/t_0^2$, $f_0=L_0\rho/t_0^2$, in which $L$ and $E$ are the length and a typical Young's modulus of the beam, respectively, the above non-dimensional \pde\ simplifies to
\begin{equation}
    \DD t\uv=\boldsymbol{\nabla} \cdot\sigma +\fv \,.\label{Eqn:StressDiv}
\end{equation}

Define the horizontal $x$-axis to be the longitudinal direction of the beam.
Along the top and bottom edges of the beam we apply stress-free boundary conditions in which stresses in the normal direction $\nv=(0,1)$ are zero:
\begin{equation}
    \sigma \cdot\nv =0,
    \quad\text{that is,}\quad 
    \sigma_{yy} = \sigma_{xy} = 0.
    \label{eqn:BCs}
\end{equation}

\subsection{Microscale spatial discretisation} 
\label{sec:discretisation}
\begin{SCfigure}
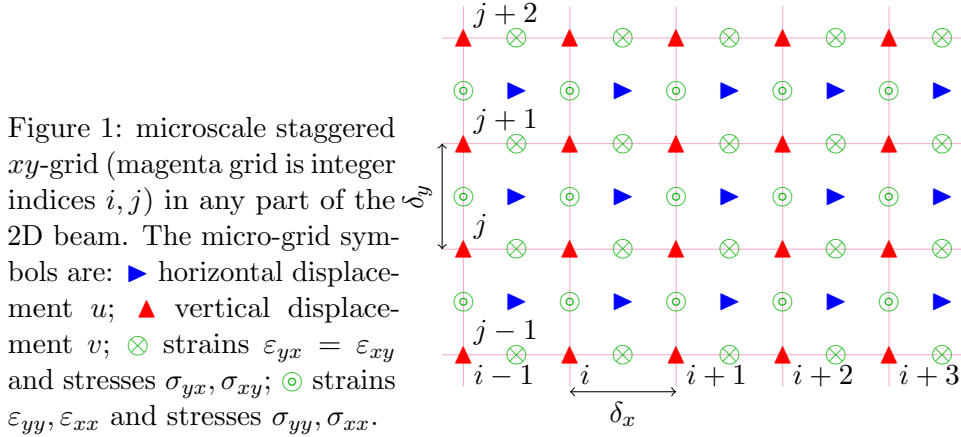

\centering
\caption{\label{microgrid}%
microscale staggered \(xy\)-grid (magenta grid is integer indices~$i,j$) in any part of the 2D~beam.
The micro-grid symbols are:
 \uSym~horizontal displacement~\(u\); \vSym~vertical displacement~\(v\);
\xSym~strains \(\varepsilon_{yx}=\varepsilon_{xy}\) and stresses~\(\sigma_{yx}, \sigma_{xy}\); 
\oSym~strains \(\varepsilon_{yy}, \varepsilon_{xx}\) and stresses~\(\sigma_{yy}, \sigma_{xx}\). }
\hspace*{-1em}\inPlot{Figs/microgrid}
\end{SCfigure}

For computational simulation we must discretise the governing microscale equations \cref{eqs:stressDef,Eqn:StressDiv} in space.
Here we invoke a straightforward microscale staggered discretisation, illustrated by \cref{microgrid}, and akin to several established schemes \citep[e.g.,][]{Virieux84, Virieux86}.

In our equation-free scheme, the microscale equations governing fields are \emph{not} modified in any way: the scheme uses whatever trusted microscale computation a user codes for numerical simulation.
Thus the following discretisation is a standard discretisation and not specific to our multiscale modelling---any valid microscale discretisation scheme is suitable, so we simply chose a common robust microscale discretisation.
\cref{sec:coupling} describes how we compute the microscale in only small sparse patches of the beam.

We choose to spatially discretise on the microscale staggered grid of \cref{microgrid}.
Microscale field values are evaluated either at integer grid nodes or at half-integer grid nodes.
Horizontal displacement~$u$ and force~$f_x$ are evaluated at half-integer nodes in both $x$~and~$y$, whereas vertical displacement~$v$ and force~$f_y$ are evaluated at integer nodes in both $x$~and~$y$.
Normal strains $\epsilon_{xx}$, $\epsilon_{yy}$ and normal stresses $\sigma_{xx}$, $\sigma_{yy}$ are evaluated at integer nodes in~$x$ but half-integer nodes in~$y$.
Shear strains~$\epsilon_{xy}$ and shear stresses~$\sigma_{xy}$ are evaluated at half-integer nodes in~$x$ but integer nodes in~$y$.

The  temporal derivatives of displacement fields~\cref{Eqn:StressDiv} in the patch interior (yellow area of \cref{figpatchgridv}) are approximated by  centred finite differences of stress fields:
\begin{subequations}\label{eqs:pdeDisc}%
\begin{align}
     \ddot{u}_{i+\half,j+\half}  & =  \frac{1}{\dx }\left(\sigma_{xx}^{i+1,j+\half}-\sigma_{xx}^{i,j+\half}\right)+\frac{1}{\dy }\left(\sigma_{xy}^{i+\half,j+1}-\sigma_{xy}^{i+\half,j}\right)+f_x^{i+\half,j+\half}, \\
    \ddot{v}_{i,k}  & =  \frac{1}{\dx }\left(\sigma_{xy}^{i+\half,k}-\sigma_{xy}^{i-\half,k}\right)+\frac{1}{\dy }\left(\sigma_{yy}^{i,k+\half}-\sigma_{yy}^{i,k-\half}\right)+f_y^{i,k},        
\end{align}
\end{subequations}
for $i=1:n_x$, $j=1:n_y-1$, and $k=1:n_y$.
Similarly, the stress fields~\cref{eqs:stressDef} in the patch are approximated by  microscale heterogeneous Lam\'e constants and centred finite differences of displacement fields:
\begin{subequations}\label{eqs:StrssEva}%
\begin{align}
    \sigma_{xx}^{i,j+\half} & =  \frac{\lambda_{i,j+\half}+2\mu_{i,j+\half}}{\dx }\left(u_{i+\half,j+\half}-u_{i-\half,j+\half}\right)
    +\frac{\lambda_{i,j+\half}}{\dy }\left(v_{i,j+1}-v_{i,j}\right),  
    \\
    \sigma_{yy}^{i,j+\half} & =  \frac{\lambda_{i,j+\half}}{\dx }\left(u_{i+\half,j+\half}-u_{i-\half,j+\half}\right)
    +\frac{\lambda_{i,j+\half}+2\mu_{i,j+\half}}{\dy }\left(v_{i,j+1}-v_{i,j}\right),  
    \\
    \sigma_{xy}^{i-\half,k} & =  \lambda_{i+\half,k}\left[\frac{1}{\dy }\left(u_{i+\half,k+\half}-u_{i+\half,k-\half}\right)
    +\frac{1}{\dx }\left(v_{i+1,k}-v_{i,k}\right)\right],    
\end{align}
\end{subequations}
for $i=1:n_x+1$, $j=1:n_y-1$, and $k=1:n_y$.

The physical elastic stress-free boundary conditions~\cref{eqn:BCs} on the top and bottom of the beam are implemented via ghost values of~$u$ and~$\sigma_{yy}$ (see \cref{figpatchgridv}).
Specifically, for $i=1:n_x+1$,
\begin{subequations}\label{eqs:bctopbot}%
\begin{align}
     \sigma_{yy}^{i,\half} &=   - \sigma_{yy}^{i,\frac32},  \quad
     \sigma_{yy}^{i,n_y+\half} = - \sigma_{yy}^{i,n_y-\half},
\\
    u_{i-\half,\half} &= u_{i-\half,\frac32} +\frac{\dy }{\dx }\left(v_{i,1}-v_{i-1,1}\right),\\
    u_{i-\half,n_y+\half} &= u_{i-\half,n_y-\half} -\frac{\dy }{\dx }\left(v_{i,n_y}-v_{i-1,n_y}\right).
    \end{align}
\end{subequations}

\paragraph{Phenomenological dissipation}
In many physical scenarios, elastic materials have weak dissipation via  internal visco-elasticity, boundary friction, and\slash or sound energy radiated to the surrounds.
To include within the scope of this exploration some \emph{representative} physical effects of dissipation, we include an additional `viscous' damping term,~$+\kappa \nabla^2\dot{\uv}$ into the non-dimensional \pde~\cref{Eqn:StressDiv}.
This damping is coded into the microscale discretisation~\cref{eqs:pdeDisc} for~\(\ddot u_{i+\half,j+\half}\)  with the additional terms 
\begin{equation}\label{eqn:viscous}
\cdots
+\frac\kappa{\dx ^2}\big(\dot u_{i-\half,j+\half}-2\dot u_{i+\half,j+\half}+\dot u_{i+\frac32,j+\half}\big)
+\frac\kappa{\dy ^2}\big(\dot u_{i+\half,j-\half}-2\dot u_{i+\half,j+\half}+\dot u_{i+\half,j+\frac32}\big),
\end{equation}
and similarly for \(\ddot v_{i,j}\).
Here we \emph{do not} justify these terms as modelling any specific physical process, they simply represent \emph{phenomenological} dissipation of `small' non-dimensional strength~\(\kappa\).

\subsection{Resolve microscale only on small sparse patches in space}
\label{sec:coupling}

The discretised visco-elastic equations~\cref{eqs:pdeDisc,eqs:StrssEva,eqs:bctopbot,eqn:viscous} of the 2D beam apply within some number~$N$ of small, well-separated, equi-sized patches of length \(h\ll L\) and width~\(W\).
These patches correspond to Representative Volume Elements in other multiscale methods.
The patches are spaced at some macroscale~\(H\) chosen to be the scale resolved by the required macroscale dynamics.
Microscale computations are only performed within the patches.

\begin{figure}
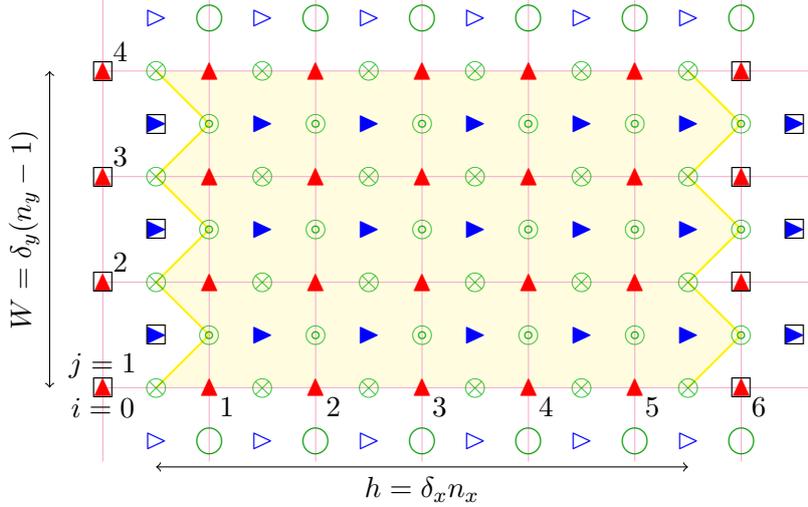

\centering
\caption{\label{figpatchgridv}%
one patch of the microscale grid of \cref{microgrid} for the case of \(n_x=5\) and \(n_y=4\) micro-grid points in each direction. 
Open symbols $\color{blue}\vartriangleright$~and~$\color{green!60!black}\bigcirc$ represent ghost nodes outside the physical beam that are used to apply the beam's top and bottom stress-free boundary conditions.  The shaded yellow region is the patch interior where field values evolve via the \pde{}s.  Symbols in squares~$\square$ (left and right) denote patch edge nodes whose values are determined from macroscale inter-patch interpolation. 
}
\inPlot{Figs/figpatchgridv}
\end{figure}%
\cref{figpatchgridv} illustrates one patch and the chosen discretisation on an equi-spaced microscale grid (magenta lines) with micro-grid spacings $\dx $~and~$\dy $ and integer grid indices $i$~and~$j$.
The \emph{interior} of each patch contains $n_x$~grid nodes in the horizontal direction and is of length \(h=(n_x-2)\dx\), with the length chosen large enough to capture the microscale structure. 
The patch also contains $n_y$~grid nodes in the vertical direction and extends across the physical width of beam $W=(n_y-1)\dy$.

\begin{figure}
\centering
\caption{\label{figpatchscheme}%
Representative volume element patches indexed by~\(I\) with a macroscale spacing~\(H\) significantly larger than the microscale patch length~\(h\). 
Each patch interior (yellow) is bounded by patch-edge values (black boxes) on both sides.
Interpolating patch interior fields (blue/red arrows) from the left/right next-to-edge values of neighbouring patches determines the edge values on the right/left of each patch~\(I\) \citep{Bunder2020a}.   
Interpolating from patches \(I-2,\ldots,I+2\) to the edges of the $I$th~patch (shown here) determines the macroscale to errors~\Ord{H^4} (e.g., see \cref{fig:Homobeam_Errors}(a)). 
}
\inPlot{Figs/figpatchscheme} 
\end{figure}%
\paragraph{Couple patches to resolve macroscale dynamics}
To capture the macroscale dynamics, we must couple the patches across unsimulated space.
While there are many ways that patches could be coupled, few couplings provide an accurate macroscale model with controllable error.
Here we apply a coupling scheme which is known to preserve self-adjoint symmetries of the microscale model~\citep{Bunder2020a}.
\cref{figpatchscheme} schematically represents this coupling, which is implemented in terms of the displacement fields~\(u,v\) near the left and right edges of each patch.
For each patch index~\(I\) and cross-beam micro-grid index~\(j\), displacement values on the right edge of the patch, \(u_{n_x+3/2,j}^I\) and~\(v_{n_x+1,j}^I\) ($\square$\,s in \cref{figpatchgridv}), are determined from interpolation of the displacement values, \(u_{3/2,j}^J\) and~\(v_{1,j}^J\) respectively, of the left next-to-edge grid-points of neighbouring patches indexed by~\(J\). 
Correspondingly, displacement fields on the left edge of a patch \(u_{1/2,j}^I\) and~\(v_{0,j}^I\) are determined from interpolation of displacement fields \(u_{n_x+1/2,j}^J\) and~\(v_{n_x,j}^J\), respectively, of the right next-to-edge grid-points of neighbouring patches~\(J\). 
These interpolated patch edge values then affect predictions of the patch scheme via the stress field equations~\cref{eqs:StrssEva}. 

An alternative inter-patch coupling is to interpolate centre-patch values in order to determine the required patch-edge values \citep{Roberts2023a}.
However, this alternative is generally not as good at preserving symmetry when applied in multiple space dimensions.

Here we apply two different types of interpolation for the patch edge fields: polynomial interpolation and spectral interpolation. 
An order~\(P\) polynomial interpolation means that the left/right edge field of the \(I\)th patch is determined from a polynomial interpolation of right/left  next-to-edge fields of patches indexed by \(I-P/2,\ldots, I+P/2\). 
Alternatively, in spectral interpolation a Fast Fourier Transform of left/right patch next-to-edge fields, via the shifting theorem, leads to an an Inverse Fourier Transform determining the right/left edge fields, respectively---this provides a global all-to-all coupling of patches in the domain. 
In general, increasing the order~\(P\) of polynomial coupling improves the accuracy of resultant macroscale predictions (\cref{fig:Homobeam_Errors}).  
Spectral coupling generally provides effectively exact macroscale predictions~\citep{Bunder2020a}.
For exploring accuracy and stability of the patch scheme applied to visco-elastic beams (\cref{Sec:Heterobeam}) we apply both spectral and polynomial interpolation.  
But we use only polynomial interpolation for predicting the equilibria and dynamics of beams with fixed \text{or free ends (\cref{sec:load}).}

\paragraph{Choose the length~\(h\) of patches}
A crucial issue in applying the patch scheme is to decide the size of the patches in space: the smaller the patches the greater the computational savings.
The patch ratio \(r:=h/H<1\) characterises the small fraction of the beam within which the microscale computations are performed.
For efficient simulations we seek to make the patch length~\(h\), and hence the ratio~$r$, as small as possible.
The smallest possible choice for~\(h\) depends on the heterogeneity, as patches must cover a statistically representative portion of the heterogeneity in order to effectively model this heterogeneity---in the case of periodic heterogeneity, \(h\) need only be as large as a single period. 
Herein we choose each patch to extend across the width of the beam (potentially one could distribute patches across the beam and further reduce computational time).  
In the case of the microscale heterogeneity having some period in space, \cite{Bunder2020a} proved controllable high accuracy is obtained by having each patch being an integral multiple of the period \cite[see also][]{Bunder2013b}.  
The smallest, most computational efficient, size is then to choose a patch length~\(h\) to encompass one microscale period \text{as done herein.}

In the case of functionally graded, near-periodic, microscale heterogeneity \cite[e.g.,][]{Anthoine2010, Shahraki2020, Shahbaziana2022}, the on-the-fly computations of the patch scheme automatically adapt to any functional graduations that occur on the macroscale. 
Consequently, in this case one chooses the patch length~\(h\) to be representative of the near-period. 
In the case of non-periodic microscale heterogeneity, \cite{Bunder2013b} discussed that best accuracy appeared to occur with core and buffer regions each about one-third of each patch, and then choose large enough patches to provide acceptable macroscale averages in the computational time available.

\subsection{Scheme is non-intrusive functional `wrapper'}
\label{secSnifw}

Recall that \cref{figpatchscheme} shows the 2D beam sampled by patches.
Then \cref{figpatchgridv} shows the zoom in to the microscale of any one patch.
This microscale is that of the \emph{given} micro-grid shown in \cref{microgrid}.
Open symbols in \cref{figpatchgridv} are ghost nodes outside the
patch and implement \emph{given} physical stress-free top\slash bottom conditions on the beam.
The \emph{only} addition required by the patch scheme are the edge values (`squared' micro-grid nodes in \cref{figpatchgridv}) on the left\slash right of each patch.

The patch scheme couples patches together by providing the patch-edge values through interpolation across the macroscale between patches \cite[e.g.,][]{Roberts06d, Roberts2011a, Cao2014a}.  
Here we interpolate from each of the \emph{next-to-edge patch values across} the beam (\(i=1,5\) in \cref{figpatchgridv}) of nearby patches, to determine the corresponding patch-edge value.
The scheme does \emph{not presume} that any particular average is appropriate.

This implementation shows that \emph{the patch scheme is non-intrusive} \cite[e.g.,][]{Biezemans2022b}: it just `wraps around' any micro-grid code a user trusts.  
Consequently, our developing toolbox \cite[]{Maclean2020a} can implement the patch scheme around micro-code for any user's particular application.

\subsection{Patch scheme embeds macroscale dynamics}
\label{SSpsemd}

Since the patch scheme does not assume any `correct' macroscale variables, \emph{a crucial question} is the following: how can we be assured that the patch scheme captures the macroscale slow dynamics?
The \cite{Whitney1936} embedding theorem provides an answer.

Roughly, the theorem establishes that every \(m\)D~manifold is parametrisable from almost every subspace of more than~\(2m\)D. 
What does this means for us?
Here, in a sense, the patch scheme provides the higher-D subspace in which the slow manifold of the macroscale visco-elastic beam dynamics is embedded.  

For beams in 2D, the basic macroscale beam models have, at each cross-section,  displacement and velocity of both bending and compression.  
Thus, at each and every cross-section, the visco-elastic beam dynamics has a basic slow manifold that is \(m=4\)D.\footnote{Such statements, invoking a manifold or subspace ``at every cross-section", are to be interpreted in the sense developed by the theory of \cite{Roberts2013a}.  That is, in systems of large spatial extent there often are important, spatially global, invariant manifolds of high-D that are effectively decomposable into a union of \emph{spatially local} manifolds\slash subspaces of relatively lower dimension---a dimension determined by the spatial cross-section---and that are weakly coupled to neighbouring locales.}
Alternatively, 2D cosserat beam models \cite[e.g.,][]{Forest2011, Somnic2022} add a rotational\slash shear mode to the macroscale model---two more variables---leading to a not-quite-so-slow manifold of \(m=6\)D at every cross-section.
These physically based models are \emph{slow manifolds} because they focus on the relatively slow visco-elastic waves varying slowly in space: they neglect all the faster high-frequency cross-waves.

\cref{sec:load,Sec:Heterobeam} discusses applications of the patch scheme to 2D beams with a cross-section of \(n_y\in\{7,\,9,\,100\}\) micro-grid points, but let's discuss the case of just \(n_y=4\) shown in \cref{figpatchgridv}.
For \(n_y=4\), there are seven microscale nodes across each patch edge.  
Each node has an associated displacement and velocity, and so leads to a \(14\)D~subspace for macroscale communication between patches.

Because \(14>2\cdot6>2\cdot4\)\,, the Whitney embedding theorem asserts that the patch scheme exchanges enough information to almost surely parametrise both such slow manifolds of the macroscale dynamics (both basic and cosserat).
The patch scheme does \emph{not} need to explicitly compute and exchange  specific assumed macroscale average quantities.

Consequently, if the microscale system actually has some unknown and/or unsuspected macroscale `average' variables, then Whitney's embedding theorem assures us that the patch scheme generally will also capture the slow dynamics of these unknown and/or unsuspected variables---just provided~\(n_y\) is not too small.

\section{Equilibria of loaded heterogeneous beams}
\label{sec:load}

This section explores the efficient calculation of equilibria of two example loaded heterogeneous beams. 
Equilibria are often the main focus of applications for homogenisation \cite[e.g.,][]{Klarmann2019, Schneider2021, Raju2021}.
Moreover, wave propagation is usually reduced to equilibria of a Helmholtz equation \cite[e.g.,][]{Rupprecht2016, Maier2020}.  
The two equilibria examples of this section utilise the flexibility of the Equation-Free Toolbox \citep{Roberts2019b}.
With these examples we explore the accuracy of the equation-free patch scheme compared to a full microscale simulation of the visco-elastic 2D beam, and extend the analysis of \cref{SSpcpn} to further test the dependence on the modelling on the number of patches~\(N\) and the implemented patch coupling.
The computational efficiency of this equation-free scheme comes from only computing the microscale heterogeneous elasticity equations on small sparse patches in space.
Simplicity in application comes from the \text{scheme being non-intrusive.}

\subsection{A fixed-fixed loaded beam}
\label{SSfflb}

This specific example is a non-dimensional beam of length~\(L=1\) and width~\(W=0.02\) that possesses a micro-structure only resolved adequately on lengths \(\dx \approx \dy \approx 0.005\) (\cref{sec:microscalemodel}). 
Here we chose the microscale Young's modulus and Poisson ratio to be random in~\([0.6,1.6]\) and~\([0.2,0.4]\), respectively, and periodic along the beam with spatial period \(\approx 0.03\) as shown in the mesh colouration of \cref{fig:equilibErrs}(a).
Here the boundary conditions at the two ends of the beam are that of fixed-ends, namely zero displacements~\(u,v\) at \(x=0,1\) for every~\(y\).
Along the top and bottom edges of the beam we apply stress-free boundary conditions. 
The thin beam is bent sideways by the non-dimensional applied force \(f_y=\frac1{1000}e^{2x}\sin x\), and we set no damping in \text{the system (\(\kappa=0\)).}

\begin{figure}\centering
\caption{\label{fig:equilibErrs}Equilibrium deflection of loaded heterogeneous elastic beam with fixed ends and width~\(0.02\) (all quantities non-dimensional).  To illustrate the heterogeneity, the mesh colour is determined by the microscale value of the Young's modulus~\(E\) (see colorbar).}
\def\extraAxisOptions{axis equal image, small, width=11cm}
\begin{tabular}{@{}l@{}}
(a) Full domain solution
\\ \inPlot{Figs/equilibErrsO6logN5}
\\ (b) \(N=17\) patches
\\ \inPlot{Figs/equilibErrsO6logN4}
\\ (c) \(N=9\) patches
\\ \inPlot{Figs/equilibErrsO6logN3}
\\ (d) \(N=5\) patches
\\ \inPlot{Figs/equilibErrsO6logN2}
\end{tabular}
\end{figure}

As in previous examples, we apply the equation-free scheme implemented by the  Equation-Free Toolbox \citep[\S3.8 and~\S4.6]{Roberts2019b}.
Spectral patch coupling is not appropriate for fixed boundary conditions, so patches are coupled via order~\(P\) polynomial interpolation.
For comparison we also compute the equilibrium for the full microscale beam (\cref{fig:equilibErrs}(a)).
The Toolbox code was used to explore the error as the number~\(N\) of patches varies in the domain, and hence as the patch spacing~\(H\) and patch ratio~\(r\) vary, whereas the patch length~\(h\) remains constant to reflect unchanging microscale physics (also similar to \cref{SSpcpn}).
\cref{fig:equilibErrs}(b)--(d) show the various predicted equilibria are graphically indistinguishable.
\cref{tbl:relerrs} summarises the errors and the compute times.
Firstly, the solution time with the patch scheme is significantly less than that for a full-domain computation---for example, five patches is here solved twenty times quicker.
In problems with greater scale separation, or in more spatial dimensions, the computational saving would be much more.

\begin{table}
\caption{\label{tbl:relerrs} the relative magnitudes of the mean-absolute errors in the simulations of the beams of \cref{fig:equilibErrs}(b)--(d), compared to the full microscale beam \cref{fig:equilibErrs}(a).  
The relative errors were computed for three different orders~\(P\) of patch coupling, and three different numbers~\(N\) of patches.}
\begin{equation*}
\begin{array}{l|cccccc}
\text{number patches } N&5&9&17&\text{full}
\\
\text{solve time (secs)}&0.2&0.5&1.1&4.3
\\\hline
\text{fourth-order }P=4&6\E-2&6\E-2&7\E-2 &0
\\
\text{sixth-order }P=6 &- &3\E-3 &1\E-3 &0
\\
\text{eighth-order }P=8 &- &1\E-4 &2\E-5 &0
\end{array}
\end{equation*}
\end{table}

Secondly, as the number~\(N\) of patches increase, the patch spacing~\(H\) decreases (\cref{fig:equilibErrs}(b--c)), and the patch scheme errors generally decrease at a rate depending upon the order of patch coupling. 
Theory by \cite{Bunder2020a} proves that for unforced problems on unbounded domains the patch scheme is consistent with the underlying microscale system to errors~\(H^P\) for order~\(P\) interpolation.
For the small range of~\(H\) (or~\(N\)) considered here, we observe that the error generally decreases with increasing~\(N\) and~\(P\), as expected.
\cref{SSpcpn} explores a wider range of patch spacings~\(H\) to clearly demonstrate the expected power law for the macroscale error.

\subsection{Comparison with some other methods}
\label{SScwsom}

\cite{Lange2023}, here often abbreviated by \lhk, recently developed a multiscale ``monolithic hyper \textsc{rom~fe}$^2$ method''\footnote{\textsc{rom} denotes Reduced-Order Modelling, and \textsc{fe}~denotes Finite Element.} for heterogeneous material, and compared it to three other extant methods in two examples.
One of the examples was the bending of a 2D heterogeneous beam (\lhk, \S5.2, Example~1).
This subsection compares our patch scheme to these previous methods as reported by \lhk\ on this example. 
The outcome is that (\cref{tbl:cfLange2023}), despite being executed on a computer with an order of magnitude less power, the patch scheme here executes in a similar time, 64~seconds in total, to that of the fastest other method, hyper-\textsc{rom}, which \lhk\ report took 56~seconds in total on their much more powerful hardware.

\begin{figure}
\centering
\caption{\label{fig:cfLange2023}%
five coupled patches model a loaded heterogeneous 2D beam of non-dimensional length~\(1\) and width~\(0.125\)---akin to Fig.~10(a) of \lhk.  The microscale heterogeneity has non-dimensional Young's modulus~\(E\) iid random in~\([0.37,2.7]\) (sub-patch colour) and Poisson's ratio~\(\nu\) iid uniform in~\([0.2,0.4]\) on a micro-grid of spacing \(\dx=\dy=0.00125\).}
\includegraphics[width=\linewidth]{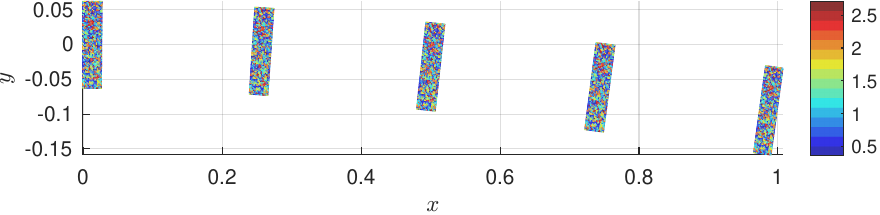}
\end{figure}

This comparison with \cite{Lange2023} (\lhk) is only indicative as there are  differences in the modelled beams and the computer hardware. 
\lhk\ (Fig.~10(a)) model a 2D beam of aspect ratio~\(8:1\) as in our \cref{fig:cfLange2023} where we set the non-dimensional length~\(1\) and width~\(0.125\).  
Fig.~10(a) of \lhk\ divides the beam into \(5\times 40\) periodic cells, each cell resolved on a micro-grid of size~\(m\times m\) for \(m=13,20,33,59\) (\lhk, Fig.~6).
Here we correspondingly choose~\(40\) periodic-cells along the beam, but each cell extends across the beam: within each cell is a \(20\times100\) micro-grid, spacing \(\dx=\dy=0.00125\)\,, in order to reasonably correspond to the five \(20\times20\) cells of \lhk.
To implement the patch scheme we choose to use five patches along the beam as shown in 
\cref{fig:cfLange2023}. 
Each patch extends across the beam, corresponding to the five cells \lhk\ use across their beam.
The patches of \cref{fig:cfLange2023} are coupled by fourth-order interpolation as indicated by \cref{figpatchscheme}.  
The beam has fixed-end boundary conditions implemented on the left (\(x=0\)) of the extreme left patch, and stress-free boundary conditions implemented on the right (\(x\approx1\)) of the extreme right patch.

For the sub-cell heterogeneity, \lhk\ use a varying Young's modulus~\(E\) of an example of \cite{Miehe2002}.  
Here we choose a roughly similar range of varying~\(E\) (colour in \cref{figpatchscheme}), but additionally vary Poisson's ratio~\(\nu\): the microscale heterogeneity has non-dimensional~\(E\) iid log-uniform random in~\([0.37,2.7]\) and~\(\nu\) iid uniform in~\([0.2,0.4]\).
These microscales are different in detail, but that is of little import as, for the same~\(m\), about the same amount of micro-scale computation is done for each cell-width across the beam.
It is because a micro-cell computation is about the same that this comparison between the algorithms may be reasonably indicative.

The ``iterative NR-loop'' of the algorithm \lhk\ depict in their Fig.~4 is replaced here by the bi-conjugate gradient algorithm (\matlab's \verb|bicgstab|) for each applied forcing (corresponding to the outer loop of Fig.~4, \lhk).
The preprocessing ``hp\textsc{rom} training'' steps in Fig.~4 of \lhk\ are here replaced by computing the Jacobian of the patch scheme wrapped around the visco-elastic patches, together with computing an incomplete \(LU\)-decomposition of the Jacobian which serves as a good preconditioner to the bi-conjugate \text{gradient algorithm.}

\begin{table}\centering
\caption{\label{tbl:cfLange2023}%
computational time (seconds), both offline and online components, from Table~1 of \protect\cite{Lange2023} (\lhk) adjoined to the patch scheme for \protect\cref{fig:cfLange2023}.
Abbreviations are: \textsc{hf}, high fidelity \textsc{fe}$^2$;
\textsc{dof}s, degrees of freedom;
\textsc{ip}s, integration points.
}\vspace*{-\baselineskip}
\begin{equation*}
\def\sv{\smash{\vdots}}
\begin{array}{|l|rrr@{\ \sv\ }r|}
\hline
&\text{two \textsc{hf}s} & \textsc{rom} & \text{hyper-\textsc{rom}} & \text{patch scheme} 
\\\hline
\text{\textsc{dof}s} & 16\,385\,400 & 14\,400 & 14\,400 & 20\,100 
\\ \text{\textsc{ip}s} & 12\,209\,400 & 12\,209\,400 & 450\,000 & 20\,100 
\\ \text{online secs} & {}\approx1100 & 215 & 23 & 12 
\\ \text{offline secs} & 0 & 12 & 33 & 52 
\\ \text{computer} & \multicolumn3{c@{\begin{array}c\sv\\\sv\\\sv\end{array}\!}}
{\parbox{15em}{16 cores of 8 Xenon(R) Gold 6244 processors, with 768\,Gb memory}} 
& \parbox{6.9em}{\raggedright MacBook Pro, 1~M2 processor, 24\,Gb memory}
\\\hline
\end{array}
\end{equation*}
\end{table}

In predicting the bending of a loaded 2D heterogeneous beam, 
Table~1 of \cite{Lange2023} compares compute times of two high fidelity \textsc{fe}$^2$ methods, and two reduced order modelling methods (\textsc{rom} and their hyper-\textsc{rom}).
Here \cref{tbl:cfLange2023} lists their characteristic side-by-side with a corresponding application of the patch scheme to much the same scenario that predicts \cref{fig:cfLange2023}.
The total compute times of the patch scheme is similar to the fastest of the other methods, despite executing on vastly less powerful computer hardware.
This indicates that the patch scheme can indeed be very effective for computing such equilibrium deformations.

Potentially, the patch scheme could achieve even higher computational efficiency by also invoking patches of cells across the beam as well as along the beam \cite[e.g.,][]{Bunder2020a}.  
But for simplicity here we use patches only along the beam.

\section{Simulate heterogeneous beam dynamics}
\label{Sec:Heterobeam}  

By far the most research and development has been done on homogenisation for problems seeking either equilibria, quasi-static laws, or a Helmholtz equation for strictly periodic linear oscillations.
However, this section addresses dynamics which is a much more stringent test of a multiscale scheme.

That dynamics are much more stringent is illustrated by at least the following two examples.
First, suppose in some scenario an analytic higher-order homogenised \pde\ is \(U_{tt}=A_2\nabla^2 U+A_4\nabla^4U+F\) for some macroscale field~\(U(\xv,t)\), homogenised coefficients~\(A_k\), and homogenised forcing~\(F\).
Then for equilibria there is rarely any difficulty in solving \(A_2\nabla^2 U+A_4\nabla^4U+F=0\)\,, neither is there any difficulty in solving the corresponding unforced Helmholtz equation \(-\omega^2U=A_2\nabla^2 U+A_4\nabla^4U\).
However, \emph{dynamically}, waves with wavenumbers \(|\kv|>\sqrt{A_2/A_4}\) are exponentially unstable and hence ruin macroscale dynamic simulations and prediction. 
Second, consider computational methodologies, like the patch scheme, that resolve both the macroscale of interest and some microscale (sub-patch) dynamics.
The behaviour of the multiscale scheme may be characterised by the eigenvalues of its linear operator for the specific given microscale.
To solve for equilibria it is almost always immaterial whether there are eigenvalues with an artificial positive real-part.
However, for dynamics, an artificial positive real-part will ruin the multiscale simulations.
Dynamics are \text{much more delicate.}

To simulate the dynamics, apply any reasonable time-integration algorithm to the system's \ode{}s~\cref{eqs:pdeDisc,eqs:StrssEva,eqs:bctopbot} within each coupled patch to calculate evolving displacements and velocities (we used \matlab's \verb|ode23|).
In this section we explore beams with macroscale periodic boundary conditions, specifically, \(2\pi\)-periodic.
Being macroscale periodic, spectral coupling may be invoked and serve as a comparison for polynomial coupling.
The functions of the equation-free toolbox \citep{Maclean2020a, Roberts2019b} `wrap around' the microscale \ode{}s to simulate the heterogeneous 2D beam.

\subsection{Simulation of beam vibrations}
\label{sec:eg}

Let's explore the example case of a heterogeneous beam of non-dimensional length~$L=2\pi$ and width~$W=0.4$, an aspect ratio of about \(16:1\). 
The heterogeneous elasticity varies at the scale of the microscale grid \(\dx=\dy=0.05\).
Specifically, the non-dimensional Young's modulus and Poisson's ratio are iid log-uniform and uniform, respectively: specifically
\begin{equation}\label{eqn:HeteElasGen}
    E(x,y)= \exp\big[\cU_1(x,y)\big],\quad
    \nu(x,y)=0.3+0.1 \cU_2(x,y),
\end{equation}
where \(\cU_k(x,y)\) are uniform on~\([-1,1]\) and iid but periodic along the beam with period~\(0.25\).
Consequently, here \(E\in[0.37,2.7]\) and \(\nu\in[0.2,0.4]\).
Patches are chosen to include one period of the microscale, and so are of size $n_x=5$ and $n_y=9$ micro-grid points in the \(x\)- and \(y\)-dimension, respectively.
Here seven patches are chosen equi-spaced in the \(2\pi\)-domain.

This example does not have a large scale separation between macro- and micro-scales---these parameters are chosen primarily so that plots visibly resolve both scales.
Nonetheless, theory guarantees the high accuracy of the macroscale predictions even in this case of a relatively small scale separation \cite[e.g.,][]{Bunder2020a}---there is no~\(\epsilon\) in the underlying theory, and no need for a `scale-separation' limit as \(\epsilon\to0\).

\begin{figure}
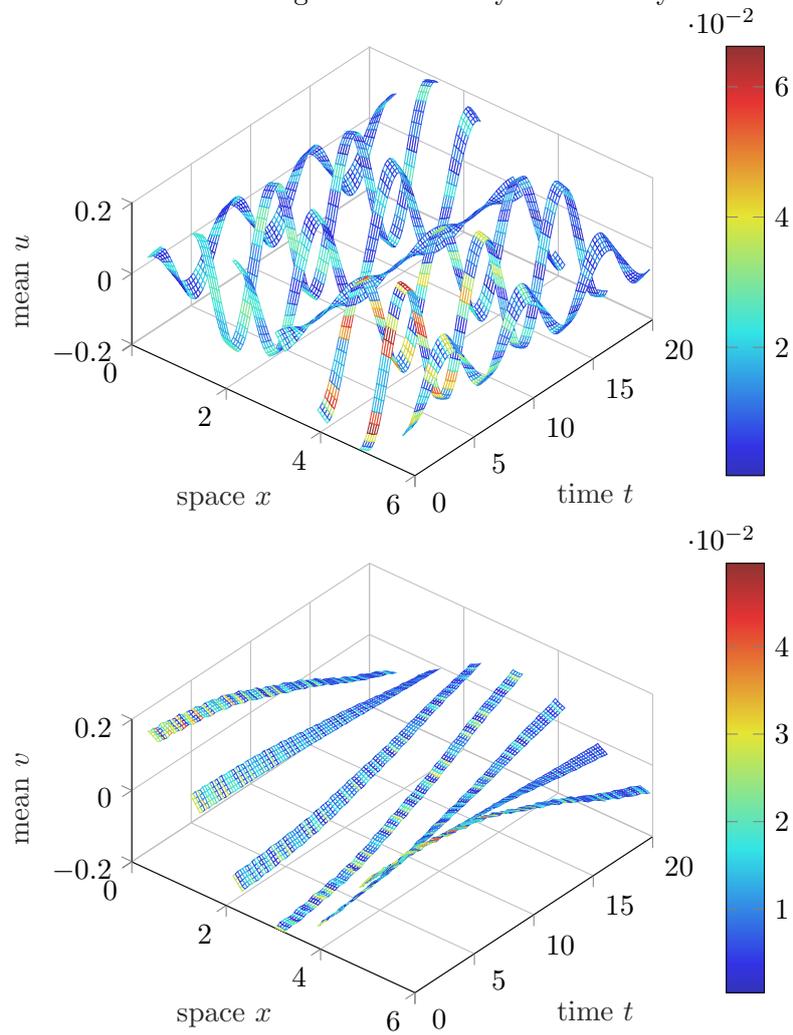
 \centering   
\caption{\label{fig:Inhomobeam_Dis}%
displacement fields of a heterogeneous beam as a function of space-time,~\(x,t\): top, mean horizontal displacement~\(\bar u\); and bottom, mean vertical displacement~\(\bar v\).
The plotted surfaces are averages over the beam's cross section in~\(y\). 
Mesh colour shows the standard deviations in displacement over each cross section. 
The heterogeneous elasticity is defined by the random~\eqref{eqn:HeteElasGen}.
}
    \inPlot{Figs/heteroPeriodicO0p7x7y9nu3u}
    \inPlot{Figs/heteroPeriodicO0p7x7y9nu3v}
\end{figure}

To illustrate the simulation of macroscale vibrations of the beam,
consider the example case of initial deformation
\begin{equation}
         u(x,y,0) =  0.2\sin{x}+0.2y\,,  \quad
         v(x,y,0)=  0.2\cos{x}+0.2y\,.
\label{Eqn:IniDeform}
\end{equation}
The \(\sin x\) in~\(u\) and \(\cos x\) in~\(v\) components excite standing waves in the beam: compression and bending waves, respectively.
The \(y\)-structure is to generate some significant out-of-quasi-equilibrium microscale dynamics which will decay due to the small viscosity \(\kappa=10^{-3}\).
The plots of \cref{fig:Inhomobeam_Dis} represent the resulting simulated evolution in space-time.

To plot the evolution of the displacements~\(u,v\) as a function of the three variables~\(x,y,t\) in all patches is hard to visualise.
At any one time instant the beam patches will look akin to \cref{fig:equilibErrs,fig:cfLange2023}, but we focus here on the evolution in time.
Consequently, at each time we average the fields across the beam in~\(y\) to obtain mean displacements~\(\bar u(x,t)\) and~\(\bar v(x,t)\) at each location within each patch.
\cref{fig:Inhomobeam_Dis} plots the surfaces of~\(\bar u(x,t),\bar v(x,t)\).
The surfaces are seven `ribbons' in the \(xt\)-space because the patch scheme only explicitly resolves the microscale within the seven patches.
\begin{itemize}
\item The sine wave of \cref{fig:Inhomobeam_Dis}(top) shows the resultant standing wave of compression waves along the beam with a non-dimensional time-period of about~\(6\).
\item The cosine wave of \cref{fig:Inhomobeam_Dis}(bottom) shows the resultant standing wave of beam bending with the longer time-period of about~\(60\).
\item The ribbon surfaces are coloured (see the colourbar) according to the standard deviation of the structure across the beam at each~\(x,t\).  
Such colouration indicates the microscale structure: here see the initial micro-structure generally decay.
\end{itemize}
The beam's simulated motions are smooth and the patch scheme does not appear to induce any instabilities in the predicted macroscale beam motions.

\cref{fig:Inhomobeam_Dis} shows just one simulation.
To understand the general solution of the patch scheme wrapped around microscale heterogeneous elasticity we analyse the whole-system dynamics via the eigenvalues and eigenvectors of its Jacobian.
We write the whole-system of patched elasticity as a system of first-ordered \ode{}s
\begin{equation}
    \dot{\qv } = J \qv \,,\label{E_JabDef}
\end{equation}
where $\qv =(u,v,\dot{u},\dot{v})$ is the generalised coordinate vector of the beam.
Computationally we determine the Jacobian by the action of the coded functions on  each of a complete set of orthonormal basis vectors for \qv-space. 
Then recall that we know a general solution for the whole patched elasticity system is of the form \(\qv=\sum_kc_k\ev_k\exp(\lambda_k t)\) in terms of eigenvalues~\(\lambda_k\) and corresponding eigenvectors~\(\ev_k\) of the Jacobian~\(J\), and for arbitrary constants~\(c_k\) depending upon the initial conditions.
Hence the set of eigenvalues capture the main information of the whole system dynamics.

\begin{SCfigure}
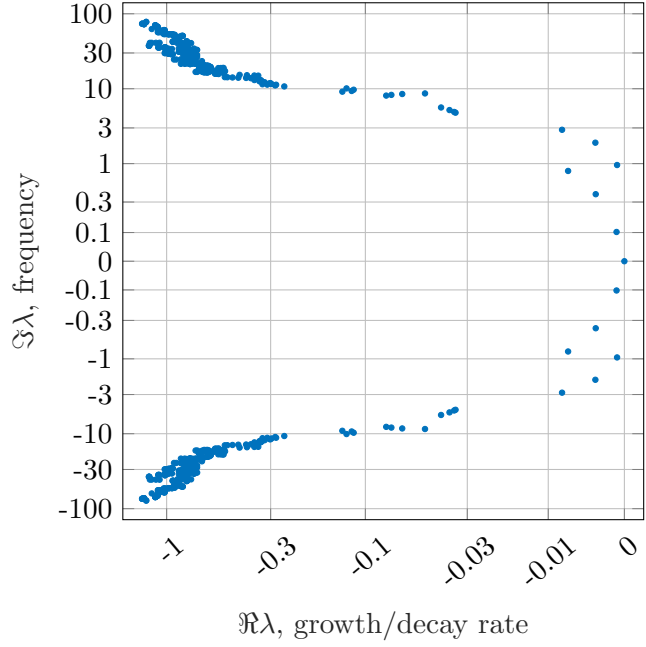
 \centering
\caption{\label{fig:Inhomobeam_EigVal}%
Eigenvalues $\lambda$ of the Jacobian in the \ode~\cref{E_JabDef} which characterise the vibration of the heterogeneous beam.
Elasticity parameters are defined by~\eqref{eqn:HeteElasGen}.
Large magnitude eigenvalues ($\Re\lambda<-0.03$, $|\Im\lambda|>3$)  correspond to inconsequential microscale sub-patch modes, and small-magnitude eigenvalues ($|\Re\lambda|<0.01$, $|\Im\lambda|<3$)  correspond to the physically interesting macroscale vibrations.}
    \inPlot{Figs/heteroPeriodicO0p7x7y9nu3Eig}
\end{SCfigure}%
\cref{fig:Inhomobeam_EigVal} plots the spectrum of all eigenvalues~$\lambda$ of Jacobian~\(J\) of the patched heterogeneous elastic beam system for the example of \cref{fig:Inhomobeam_Dis}.
\cref{fig:Inhomobeam_EigVal} is typical and does not differ significantly for other examples.

\begin{SCfigure}
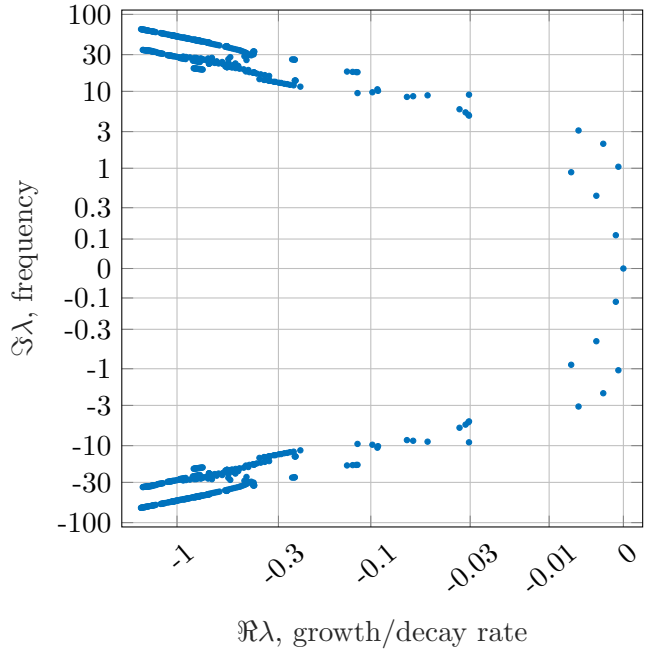
 \centering
\caption{\label{fig:Homobeam_EigVal}%
Eigenvalues \(\lambda\) of the Jacobian for the homogeneous beam with constant \(E=1\) and \(\nu=0.3\).
The spectral structure of eigenvalues is much the same as for the heterogeneous case shown in \protect\cref{fig:Inhomobeam_EigVal}. 
Some differences occur in the large magnitude eigenvalues ($\Re\lambda<-0.03$ and $|\Im\lambda|>3$) due to the differences in the microscale elasticity.}
    \inPlot{Figs/homoPeriodicO0p7x7y9nu3Eig}
\end{SCfigure}

\paragraph{Stability}
Observe in \cref{fig:Inhomobeam_EigVal}, and also in \cref{fig:Homobeam_EigVal,fig:HeteroNoVisEigVal,fig:Inclusionbeam_EigVal}, that there are \emph{no} eigenvalues with positive real-part.
This reflects that the patch scheme preserves the stability of the underlying heterogeneous visco-elasticity.

\paragraph{Dynamics}
We classify the eigenvalues in spectra such as \cref{fig:Inhomobeam_EigVal} into two main groups: those with large magnitude (to the left in the plot); and those with small magnitudes (to the right in the plot).
\begin{itemize}
\item The large magnitude eigenvalues in \cref{fig:Inhomobeam_EigVal},  frequencies \(|\Im\lambda|>3\), represent small sub-patch waves that dissipate relatively quickly, here with decay rates \(-\Re\lambda>-0.03\).
These sub-patch waves `bounce' around within a patch, albeit leaking via the coupling to other patches, as they decay. 
They make no contribution to the macroscale predictions of interest. 
 
\item The macroscale waves of interest are represented by the eigenvalues of relative small frequencies in \cref{fig:Inhomobeam_EigVal},  frequencies \(|\Im\lambda|<3\), that dissipate slowly, here the decay \(-\Re\lambda<0.01\). 
Let's physically interpret these eigenvalues.
\begin{itemize}
\item There are four eigenvalues \(\lambda=0\) (to within round-off error) representing the rigid body motions of the \(2\pi\)-periodic beam.

\item One of two branches of eigenvalues are formed by the three pairs of complex conjugate eigenvalues  \(-0.001 \pm\I 0.958\), \(-0.003 \pm\I 1.906\), \(-0.008 \pm\I 2.833\) (each of multiplicity two).  
These reflect macroscale compression waves (\cref{fig:Inhomobeam_Dis}, top) along the beam for the three macroscale wavenumbers resolvable by just seven patches.

\item The other branch of eigenvalues are formed by the three pairs of complex conjugate eigenvalues  \(-0.001 \pm\I 0.102\), \(-0.003 \pm\I 0.385\), \(-0.007 \pm\I 0.797\) (each of multiplicity two).  
These reflect the lower frequency, longer period, macroscale beam bending waves (\cref{fig:Inhomobeam_Dis}, bottom) for the three macroscale wavenumbers resolvable by the seven patches.
\end{itemize}
\end{itemize}
The gap in the eigenvalues over \(-0.03<\Re\lambda<-0.01\) is due to these mid-range eigenvalues not being supported across the unsimulated space between patches. 
A more practical smaller patch length~\(h\), relative to the patch spacing~\(H\), would increase the size of this gap in the eigenvalue spectrum (e.g., the case of \cref{fig:Inclusionbeam_EigVal}(a)).

\paragraph{Macroscale effect of heterogeneity}
As a qualitative test of the computational homogenisation provided by the patch scheme, we compare its spectrum with that of a corresponding homogeneous beam.
For the beam of \cref{fig:Inhomobeam_Dis,fig:Inhomobeam_EigVal} the average (non-dimensional) Young's modulus and Poisson's ratio are close to~\(1\) and~\(0.3\), respectively.
So here we compare with corresponding results for a beam with constant \(E=1\)\,, \(\nu=0.3\)\,, and all other parameters the same.
Simulations of a \emph{homogenous} beam with these constant values of Young's modulus and Poisson's ratio produce mean horizontal and vertical displacements visually identical to those of the heterogeneous simulation (\cref{fig:Inhomobeam_Dis}).
\cref{fig:Homobeam_EigVal} plots all eigenvalues of the homogeneous Jacobian, and the general pattern is very similar to those of the heterogeneous case seen in \cref{fig:Inhomobeam_EigVal}.
Some differences occur in the microscale eigenvalues (to the left in the plot), as expected, because the homogeneous and heterogeneous beams have different microscale structure. 
However, there is no noticeable difference in the macroscale eigenvalues, so the macroscale predictions \text{are much the same.}

\subsection{Patch scheme has controllable error}
\label{SSpcpn}

An advantage of the equation-free patch scheme is that computational costs are reduced by only simulating within small patches of length \(h\ll L\), and thus only a small fraction~\(r\) of the total domain is simulated. 
A simulation with fewer patches requires fewer calculations, but too few patches would fail to adequately resolve the desired macroscale features in the simulation.
Simulation accuracy depends on both the chosen inter-patch coupling and the patch spacing~\(H\).

Often we couple patches with spectral interpolation as this provides effectively exact macroscale predictions for the scales resolved by the patch distribution~\citep{Bunder2020a}. 
For example, here we find that with spectral interpolation the eigenvalues of macroscale bending and compression waves are constant, to within a round-off error~\(10^{-11}\), independent of the number of patches~\(N\).
This section investigates the accuracy of different orders of polynomial interpolation, \(P\in\{4,6,8\}\), as measured by the relative error compared to the effectively exact spectral interpolation.
For each order, we determine the dependence of the accuracy on the number of patches~\(N\), and hence upon their spacing as \(H\propto 1/N\). 
As we are interested in macroscale predictions, we determine the accuracy of a particular interpolation scheme from the eigenvalues corresponding to the macroscale bending \text{and compression waves.}

For each order~\(P\) of polynomial interpolation, the number of patches~\(N\) is varied from~$5$ to $10$, $20$ and~$40$, while the beam length \(L=2\pi\)\,, beam width \(W=0.2\)\,, patch length~\(h=\pi/40\), viscosity \(\nu=10^{-3}\), and the microscale spatial resolution~\((\dx ,\dy )\) are fixed. 
As the number of patches increase, the patch spacing~\(H\) decreases, and the patch ratio~\(r\) increases from~$0.0625$ to~$0.125$, $0.25$ and~$0.5$, respectively. 
The computational costs double with each doubling of~\(N\)  (equivalently~\(r\)).
But as \(N\)~doubles, the length scales resolved on the macroscale double and so we expect the accuracy to increase.

\begin{figure}
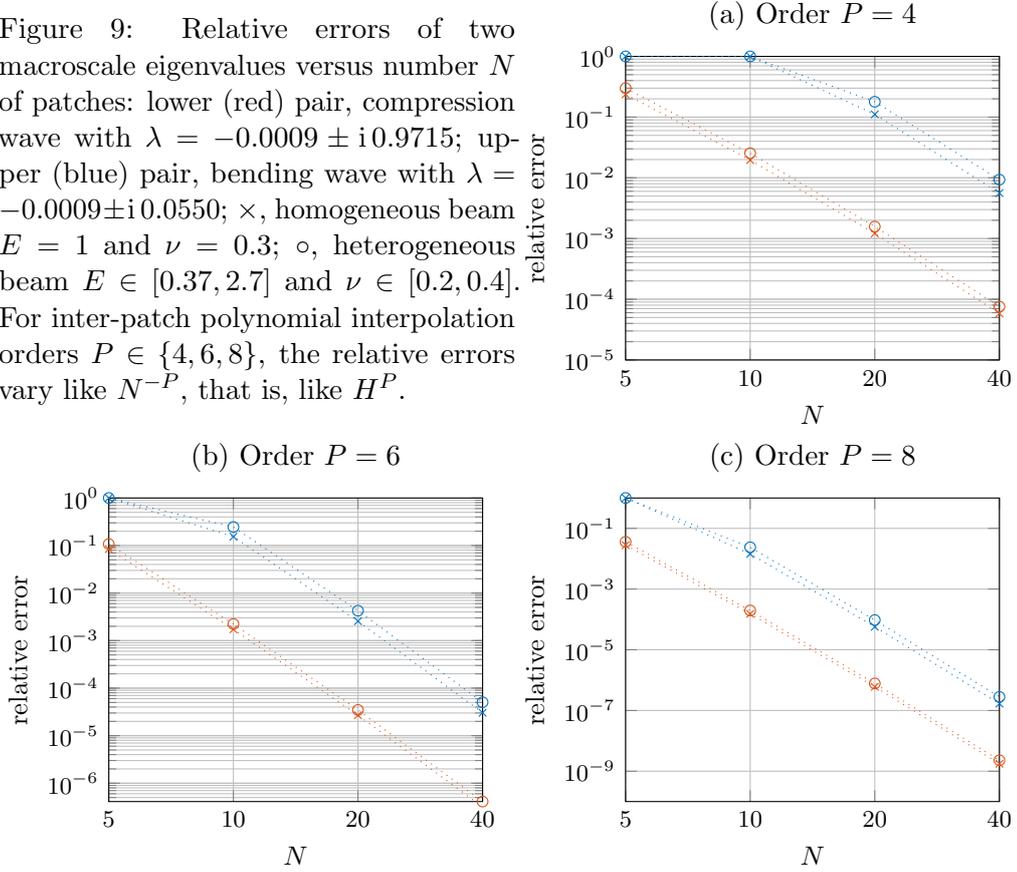
\centering
\begin{tabular}{@{}l@{}l@{}}
\parbox[t]{68mm}{\caption{\label{fig:Homobeam_Errors}%
Relative errors of two macroscale eigenvalues versus number~\(N\) of patches:  
lower (red) pair, compression wave with \(\lambda=-0.0009 \pm\I 0.9715\);
upper (blue) pair, bending wave with \(\lambda=-0.0009 \pm\I 0.0550\);
$\times$, homogeneous beam \(E=1\) and \(\nu=0.3\);
$\circ$, heterogeneous beam \(E\in[0.37,2.7]\) and \(\nu\in[0.2,0.4]\).
For inter-patch polynomial interpolation orders \(P\in\{4,6,8\}\),  
the relative errors vary like~\(N^{-P}\), that is, like~\(H^P\).}}
&   \raisebox{-\height}{\inPlot{Figs/errorOrderx7y9kap3O4copy}}
\\  \inPlot{Figs/errorOrderx7y9kap3O6copy}
&   \inPlot{Figs/errorOrderx7y9kap3O8copy}
\end{tabular} 
\end{figure}%
\cref{fig:Homobeam_Errors} plots the relative errors of eigenvalues corresponding to bending (blue) and compression (red) macroscale waves. 
That the relative errors of the bending waves are larger than those of the compression waves is largely due to the frequency of bending being twenty times smaller than that of compression.
The three subplots are for different orders of polynomial interpolation.
From the different scalings of the vertical axis, observe that the relative errors roughly like~\(N^{-P}\), that is~\(H^P\).
This power-law dependence is predicted by the theory of \cite{Bunder2020a}.

\cref{fig:Homobeam_Errors} plots the errors for two different cases of microscale elasticity.  
The crosses~\(\times\) are for a homogeneous beam with \(E=1,\,\nu=0.3\), whereas the circles~\(\circ\) are for a realisation of iid random heterogeneity~\eqref{eqn:HeteElasGen}.
The small differences in the errors between these two cases appear insignificant.
That is, and in agreement with theory \citep{Bunder2020a}, the patch scheme is just as accurate for heterogeneous elasticity as it is \text{for homogeneous.}

\subsection{Elasticity simulations without damping}
\label{SSeswd}

In the simulation of the heterogeneous beam a phenomenological damping, via a viscosity, dissipates high-frequency modes, and so smoothes the beam's motions  (e.g., \cref{fig:Inhomobeam_Dis}). 
Such dissipation also weakly dampens the beam's macroscale modes, as evidenced by the small non-zero real parts of the eigenvalues. 
However, many scientists and engineers will also want to simulate undamped motion.
In such scenarios, the slightest defect in a multiscale methodology may induce  artificial exponentially growing instabilities \text{that ruin simulations.}

\begin{SCfigure} \centering    
\caption{\label{fig:HeteroNoVisEigVal}%
Eigenvalues~\(\lambda\) of the patch scheme applied to an example of an undamped heterogeneous beam.
Eigenvalues with large magnitude ($|\Im\lambda|>9$) correspond to physically inconsequential microscale sub-patch modes. 
Whereas small magnitude eigenvalues ($|\Im\lambda|<3$) correspond to  macroscale waves of physical interest.
All eigenvalues have zero real-part, to within round-off error, to accurately reflect stable wave propagation predicted via the patch scheme (see \cref{fig:HeteroNovis_Dis}).}
    \inPlot{Figs/heteroPeriodicO0p7x7y9kap0Eig}
\end{SCfigure}%
By design, the patch scheme implemented herein preserves important symmetries of many physical systems \citep{Bunder2020a}.\footnote{The \matlab\ toolbox implements the symmetry preserving patch scheme when invoked by specifying \emph{EdgyInt} interpolation \citep{Maclean2020a, Roberts2019b}.}
Consequently, the particular patch scheme described herein does \emph{not} induce artificial instabilities.
\cref{fig:HeteroNoVisEigVal} plots the eigenvalues for a specific example of the patch scheme applied to an undamped heterogeneous beam (\(\kappa=0\)).
All eigenvalues have zero real-part (to within round-off error): the pair of eigenvalues of order~\(\pm10^{-6}\) are due to the zero-eigenvalue being of multiplicity four, and hence being very sensitive to round-off.
Consequently, it appears this patch scheme can be used to reliably simulate the dynamics \text{of undamped elasticity.}

\begin{figure}
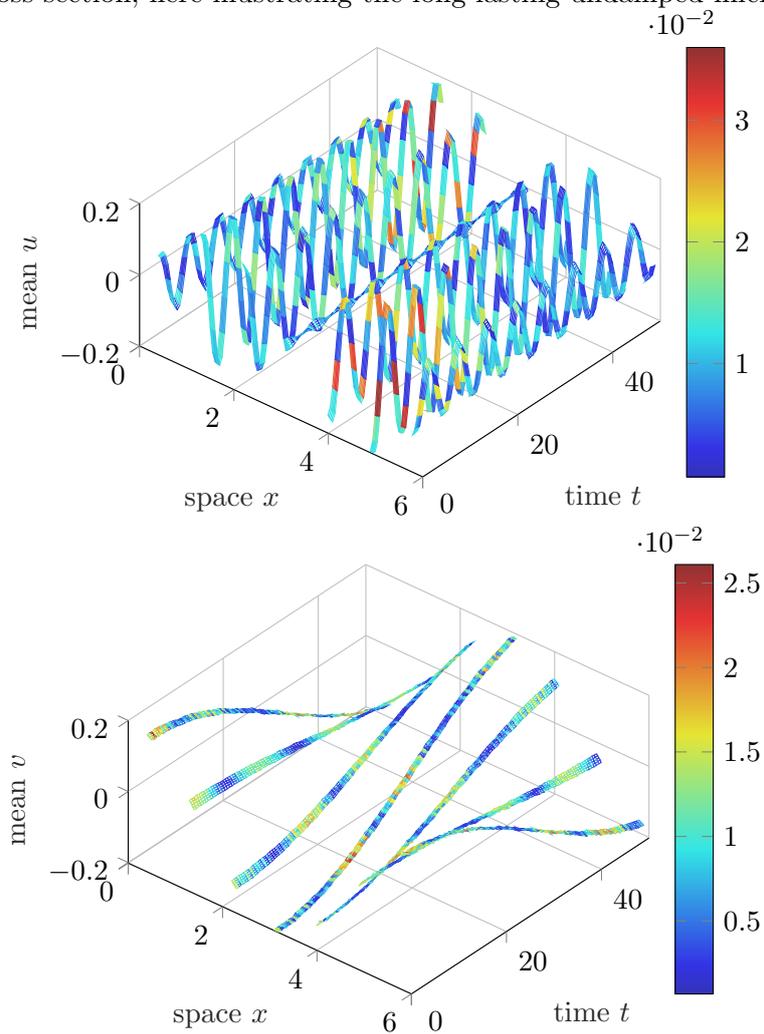
 \centering   
\caption{\label{fig:HeteroNovis_Dis}%
displacement fields of an undamped heterogeneous beam as a function of space-time,~\(x,t\), where the heterogeneous elasticity is the random~\eqref{eqn:HeteElasGen}: 
top, mean horizontal displacement~\(\bar u\); and 
bottom, mean vertical displacement~\(\bar v\).
The plotted surfaces are averages over the beam's cross section in~\(y\). 
Mesh colour shows the standard deviations in displacement over each cross section, here illustrating the long-lasting undamped microscale.}
    \inPlot{Figs/heteroPeriodicO0p7x7y9kap0u}
    \inPlot{Figs/heteroPeriodicO0p7x7y9kap0v}
\end{figure}
\cref{fig:HeteroNovis_Dis} plots the horizontal and vertical motions of a heterogeneous undamped beam as predicted by the patch scheme.  
Compared with \cref{fig:Inhomobeam_Dis}, here the beam is thinner, \(W=0.2\), has smaller microscale \(\dx=\dy=0.025\), and is resolved on smaller patches.   
The thinner beam means that the bending wave has a longer period, here about~\(100\), whereas the compression wave has much the same period.
Without damping, microscale fluctuations are observable in the displacements throughout the simulation by the colouration of the mesh that encodes the standard deviation of the displacements over the cross sections of the beam.
Undamped simulations like \cref{fig:HeteroNovis_Dis} would continue to show significant microscale fluctuations for all time.

\subsection{Beam with microscale soft inclusions}
\label{Sec:SoftInclusion}

Previous examples of this article consider beams with random microscale heterogeneity in the elasticity.
However, synthetic composite materials often have a regular microscale heterogenous structure, such as multiple layers of materials with distinctly different elastic properties.
Here we consider the specific example of a composite beam which is predominantly made of a homogenous elastic material, non-dimensionally \(E=1\) and \(\nu=0.3\), but with imbedded, identical inclusions of a `soft' homogeneous material, for various Young's modulus~\(E_\text{in}\), regularly spaced with microscale spacing~\(h\). 
To apply the patch scheme, we choose the patches to be of length~\(h\) so that each patch covers \text{one microscale period.}

For a definite example, we set the inclusions to be rectangular with length~$l$ and width~$w$, and positioned in the centre of the beam's cross section.
Each patch is placed to centre on an inclusion.
\cref{fig: Inclusion} illustrates a patch with an inclusion of length $l=4\dx$ and width $w=2\dy$\,.
\begin{SCfigure}
\centering
\caption{\label{fig: Inclusion}%
Example patch of \(n_{x}=9\), \(n_y=7\) (pale yellow) of the microscale grid with a soft inclusion (pale blue) in the beam. 
Each inclusion has length \(l=4\dx \) and width \(w=2\dy\).
Each patch is centred on one of the regularly spaced soft inclusions.}
\hspace*{-1.8em}\inPlot{Figs/figinclusion}
\end{SCfigure}

\begin{figure} \centering   
\caption{\label{fig:Inclusionbeam_Dis}%
Mean displacements of the homogeneous beam with soft inclusions versus longitudinal and time dimension: top, the horizontal displacement; and bottom,  the vertical displacement over a longer time.}
\inPlot{Figs/Inclusionbeam_Meanu}
\inPlot{Figs/Inclusionbeam_Meanv}
\end{figure}%
In the case of the inclusion with Young's modulus $E_\text{in}=0.1$ (one tenth of the main beam material) and the same Poisson's ratio \(\nu_\text{in}=0.3\),
\cref{fig:Inclusionbeam_Dis} plots the mean displacements of the composite beam  from the initial deformation~\cref{Eqn:IniDeform}.
Observe small microscale out-of-quasi-equilibrium vibrations in the very early stages of the simulation, especially in the bending wave. 
These microscale vibrations occur because the smooth initial deformation~\cref{Eqn:IniDeform} is generally \emph{not} in quasi-equilibrium: they are more noticeable here because the structure of the inclusion is more coherent on the microscale than the iid random heterogeneity~\cref{eqn:HeteElasGen} of the previous examples.
The coherent interfaces between the main beam material and the inclusions here generate more coherent microscale transients that are observable in the simulations.
The small viscosity of \(\kappa=10^{-3}\) soon dissipates \text{such microscale dynamics.}

\begin{table}
\caption{Eigenvalues for the compression and bending modes of a composite beam with soft inclusions with different Young's modulus~\(E_\text{in}\).}
    \label{tab:Inclusionbeam_Eig}
\begin{equation*}%
    \begin{array}{lcc}
       \text{Young's modulus}  &  \multicolumn{2}{c}{\text{Eigenvalues}}  \\       
       \text{of inclusions}~E_\text{in}  & \text{Compression} & \text{Bending}\\
       \hline
       1.  & -0.001 \pm\I 1.048& -0.001 \pm\I 0.059 \\
       0.1 &  -0.001 \pm\I 0.909 & -0.001 \pm\I 0.056 \\
       0.01 &  -0.002 \pm\I 0.859 & -0.001 \pm\I 0.055\\
       0.001 &  -0.014 \pm\I 0.850 & -0.001 \pm\I 0.054 \\ 
       0.0001 & -0.014 \pm\I 0.862 & -0.001 \pm\I 0.054\\\hline
    \end{array}
\end{equation*}
\end{table}%
From the simulation of \cref{fig:Inclusionbeam_Dis} the bending and compression wave frequencies are measured at~$0.056$ and~$0.909$, respectively, and are about~$4.5\%$ and~$13.2\%$ less than the frequencies of the homogeneous beam (without inclusions).
As expected, the beam is made less elastic by introducing such soft inclusions: the inclusion affects the compression waves more than the bending waves. 
\cref{tab:Inclusionbeam_Eig} lists the compression and bending frequencies for inclusions with a range of different Young's modulus.
The first case in this table has \(E_{\text{in}}=1.0\) and is a homogeneous beam (the inclusions are identical to the main beam material), whereas the following four cases have inclusions of `softer' material, lower~\(E_{\text{in}}\). 
A general decrease in frequencies (imaginary part of eigenvalue) is observed as \(E_{\text{in}}\) is decreased, with the exception of the smallest considered case \(E_{\text{in}}=0.0001\)\,.
The soft inclusions also increase the damping of compression waves, as shown by the increased magnitude of the real part of the eigenvalue as the inclusion Young's modulus is decreased.

\begin{figure}
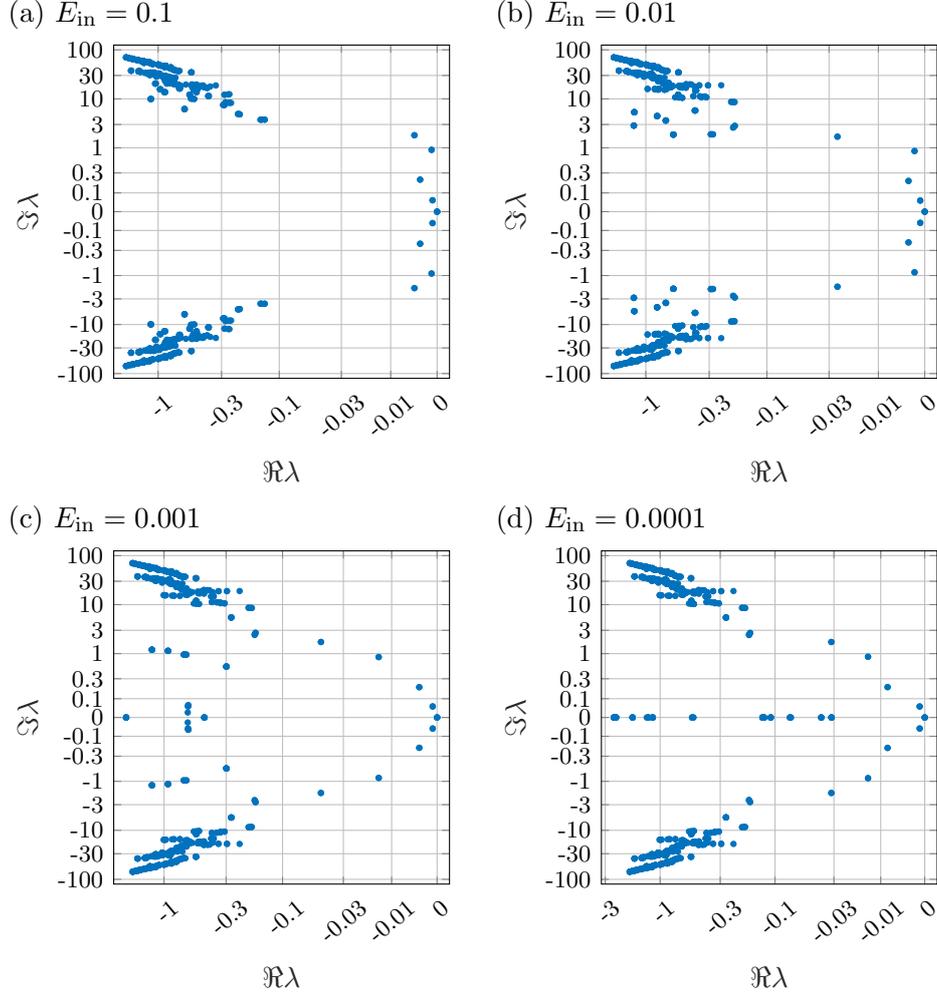
 \centering
\caption{\label{fig:Inclusionbeam_EigVal}%
Eigenvalues \(\lambda\) of the Jacobian of four composite beams with inclusions of different Young's modulus~$E_\text{in}$.  
The main beam material has Young's modulus  $E=1$ and Poisson's ratio $\nu=0.3$\,.}
\begin{tabular}{@{}ll@{}}
(a) $E_\text{in}=0.1$ & (b) $E_\text{in}=0.01$ \\
\inPlot{Figs/Inclusionbeam01_Eigval} &
\inPlot{Figs/Inclusionbeam001_Eigval} \\
(c) $E_\text{in}=0.001$ & (d) $E_\text{in}=0.0001$ \\
\inPlot{Figs/Inclusionbeam0001_Eigval} &
\inPlot{Figs/Inclusionbeam00001_Eigval}
\end{tabular}
\end{figure}%
\cref{fig:Inclusionbeam_EigVal} plots the complete eigenvalue spectra of the four composite beams reported in \cref{tab:Inclusionbeam_Eig}. 
As the inclusion Young's modulus~\(E_{\text{in}}\) is decreased, the macroscale compression wave eigenvalues shift to the left as their real-part decreases, and thus are increasingly damped (as expected from \cref{tab:Inclusionbeam_Eig}), although the bending modes are not significantly affected.
Some microscale eigenvalues are also affected by the inclusions, with their imaginary part decreasing as the inclusion Young's modulus  decreases: these eigenvalues correspond to vibrational modes within the inclusions which have smaller frequencies due to their weaker elastic properties. 
For the smallest inclusion Young's modulus \(E_{\text{in}}=0.0001\) there are several microscale eigenvalues with zero imaginary part, characteristic of overdamped modes within the inclusions that are essentially inelastic.
The remaining microscale modes evident in \cref{fig:Inclusionbeam_EigVal} represent sub-patch vibrations within the \text{main elastic material.}

\section{Conclusion}
\label{Sec:Conclusion}

Modern composite materials and structures are inherently heterogeneous across multiple scales.  
Multiscale modelling empowers us to predict the macroscale characteristics from detailed microscale material properties, and, for example, is critical to the optimal design of composite structures.
Here we have begun to develop the Equation-Free Patch Scheme as a practical multiscale method for visco-elasticity by applying it to heterogeneous 2D beams.
Future research will develop the patch scheme to 3D visco-elastic beams and shells, with anisotropic and/or nonlinear microscales.

\cref{Sec:Patchscheme} introduced one way to implement the multiscale patch scheme on the given microscale visco-elasticity equations, to empower efficient and accurate macroscale predictions of the beams.
Importantly, the scheme is non-intrusive (\cref{secSnifw}), and established theory assures us that it will reasonably capture all necessary macroscale `variables' (\cref{SSpsemd}).

Equilibria are often the main focus of applications for homogenisation, and wave propagation is often reduced to finding equilibria of a Helmholtz equation, so \cref{sec:load} explored the characteristics of the patch scheme for predicting equilibria of loaded heterogeneous beams.  
For a fixed-fixed beam, \cref{SSfflb} showed convergence of the patch scheme towards the exact, fully detailed, equilibrium as one increases the number of patches and/or the order of inter-patch interpolation.
For a fixed-free beam, \cref{SScwsom} discussed how the patch scheme is comparably computationally efficient as the best of some other \text{established multiscale schemes.}

But dynamics are a much more stringent test of a multiscale method.
Especially in common scenarios, like elastic structures, where the dissipation is weak, so the system is `close' to marginal stability, and the slightest numerical artifice may tip the multiscale method into ruinous instability.
\cref{Sec:Heterobeam} establishes that the multiscale patch scheme developed here is numerically stable---even for the case of no dissipation (\cref{SSeswd}).
Further, \cref{SSpcpn} shows that the patch scheme has controllable error for predicting the macroscale homogenised dynamics of microscale heterogeneity: by varying patch spacing and/or order of inter-patch interpolation one may straightforwardly predict macroscale characteristics to \text{eight significant digits!}

These results indicate that the multiscale patch scheme developed here forms a sound, flexible, and efficient basis for predicting the behaviour of modern `smart' heterogeneous materials. 
Scientists and engineers can start using the scheme on physical problems of their interest via the freely-available toolbox \citep{Roberts2019b}.

\paragraph{Acknowledgements}
This research was supported by Australian Research Council grants DP220103156 and DP200103097.

\end{document}